\documentclass[aps,prd,fleqn,floatfix,nofootinbib,showpacs,twocolumnpreprintnumbers]{revtex4}
\usepackage{amsmath,amssymb,amsfonts}
\usepackage{graphicx}
\usepackage{bm}
\usepackage{color}

\begin{document}

\title{Nuclear dependence of azimuthal asymmetry in semi-inclusive deep inelastic scattering}
\author{Jian-Hua Gao$^{1,2,3}$ Zuo-tang Liang$^2$, Xin-Nian Wang$^{3}$}
\address{$^{1}$Department of Modern Physics, University of Science and Technology of China, Hefei, Anhui 230026, China}
\address{$^2$Department of Physics, Shandong University, Jinan, Shandong 250100, China}
\address{$^3$Nuclear Science Division, MS 70R0319, Lawrence Berkeley National Laboratory, Berkeley, California 94720}

\date{\today}

\preprint{LBNL-}


\begin{abstract}
Within the framework of a generalized factorization, semi-inclusive deeply inelastic scattering (SIDIS) cross
sections can be expressed as a series of products of collinear hard parts and transverse-momentum-dependent (TMD) 
parton distributions and correlations. The azimuthal asymmetry
$\langle\cos\phi\rangle$ of unpolarized SIDIS in the small transverse momentum region will depend on both
twist-2 and 3 TMD quark distributions in target nucleons or nuclei. Nuclear broadening of these twist-2 and 3 quark
distributions due to final-state multiple scattering in nuclei is investigated and the nuclear dependence of the
azimuthal asymmetry $\langle\cos\phi\rangle$ is studied. It is shown that the azimuthal asymmetry is suppressed
by multiple parton scattering and the transverse momentum dependence of the suppression depends on the
relative shape of the twist-2 and 3 quark distributions in the nucleon. A Gaussian ansatz for TMD  twist-2 and 3
quark distributions in nucleon is used to demonstrate the nuclear dependence of the azimuthal asymmetry
and to estimate the smearing effect due to fragmentation.
\end{abstract}


\maketitle


\section{Introduction}

In high-energy nuclear collisions, from deeply inelastic
scattering (DIS) off a nuclear target to hadron-nucleus and
nucleus-nucleus collisions, multiple parton scattering plays
an important role in the reaction dynamics and the final hadron
spectra.  It not only causes the transverse momentum broadening of
the propagating parton but also leads to parton energy loss due to
induced gluon bremsstrahlung. A direct consequence of the
transverse momentum broadening is the so-called Cronin effect or
the broadening of the final hadron spectra in transverse momentum
in both  DIS off nuclear targets
\cite{VanHaarlem:2007kj,VanHaarlem:2009zz} and
hadron-nucleus collisions \cite{Cronin:1974zm,Antreasyan:1978cw}.
Gluon bremsstrahlung induced by multiple parton scattering on the
other hand leads to parton energy loss \cite{Gyulassy:1993hr,Baier:1996sk,
Wiedemann:2000za,Gyulassy:2000er} and
medium modification of the parton fragmentation functions
\cite{Guo:2000nz,Wang:2001ifa}, a phenomenon  known as jet quenching which is observed as the
suppression of the leading hadron yields from parton
fragmentation. The radiative parton energy loss and medium
modification to parton fragmentation functions are also determined
by the jet transport parameter or the average squared transverse
momentum broadening per unit distance experienced by the
propagating parton, which is related to the gluonic structure of
the scattering centers in the medium \cite{Baier:1996sk,CasalderreySolana:2007sw}. Therefore,
extraction of the jet transport parameter through experimental
measurements of transverse momentum broadening and jet quenching
can provide important information about the properties of the
medium, either cold nuclei or hot and dense QCD matter, as probed
by energetic partons. The jet quenching phenomenon has been seen
in high-energy heavy-ion collisions via strong suppression of not
only the large transverse momentum single hadron spectra
\cite{Adcox:2001jp,Adler:2002xw} but also the back-to-back dihadron
correlation \cite{Adler:2002tq}. The extracted jet transport
parameter is found to be much larger than that in large cold
nuclei as obtained by phenomenological studies of jet quenching in DIS
\cite{Wang:2002ri,Bass:2008rv}, indicating much higher gluonic density in the initial
stage of high-energy heavy-ion collisions.


Much efforts have also been devoted to the study of transverse momentum broadening in
DIS off nuclear targets \cite{Dolejsi:1993iw,Johnson:2000dm,Baier:1996sk,
Luo:1992fz,Guo:1998rd,Osborne:2002st,Wang:2001ifa,Guo:2000nz, Raufeisen:2003zk,Liang:2008vz}
and the Drell-Yan dilepton production in proton-nucleus collisions \cite{Bodwin:1988fs,Michael:1981yy,Guo:1999eh}
 within different approaches such as
the color dipole model\cite{Dolejsi:1993iw,Johnson:2000dm,Domdey:2008aq} and
higher-twist expansion in the generalized collinear factorization formalism  \cite{Guo:1998rd,Luo:1992fz}.
It has been shown recently \cite{Liang:2008vz} that the gauge invariant transverse momentum
dependent (TMD) quark distributions in nucleons and nuclei can be expressed as a sum of higher-twist
collinear parton matrix elements,
\begin{eqnarray}
f_q^A(x, k_\perp)&=&\int \frac{dy^-}{2\pi}e^{ixp^+y^-} 
\langle A \mid \bar\psi(0)\frac{\gamma^+}{2}
e^{\vec W_\perp(y^-) \cdot\vec\nabla_{k_\perp}}
\psi(y^-)\mid A \rangle
\delta^{(2)}(\vec k_\perp),
\end{eqnarray}
in terms of the parton transport operator
\begin{eqnarray}
\vec W_\perp(y^-)&\equiv& i\vec D_\perp(y^-)
+g\int_{-\infty}^{y^-}d\xi^- \vec F_{+\perp}(\xi^-),
\label{eq:transop}
\end{eqnarray}
where $\vec D_\perp(y^-)=\vec\partial_\perp+ig\vec A_\perp(y^-)$ is the covariant derivative.
For brevity of presentation, we have used the light-cone gauge and all the transverse coordinates
have been set to $\vec y_\perp=\vec 0_\perp$ in the above. Since the above expression is valid for both
nucleon and nuclear targets, the nuclear dependence of TMD parton distribution functions
will come from the nuclear dependence of the higher-twist parton matrix elements.
Under the ``maximal two gluon approximation", the TMD quark distribution in a nucleus
can be expressed  as a convolution of the TMD quark distribution $f_q^N(x,\vec\ell_\perp)$ in nucleon
and a Gaussian broadening \cite{Liang:2008vz} ,
\begin{equation}
f_q^A(x,k_\perp)\approx\frac{A}{\pi \Delta_{2F}}
\int d^2\ell_\perp e^{-(\vec k_\perp -\vec \ell_\perp)^2/\Delta_{2F}}f_q^N(x,\ell_\perp).
\label{eq:tmdNA}
\end{equation}
The broadening width $\Delta_{2F}$ or the total average squared transverse momentum broadening,
\begin{equation}
\Delta_{2F}=\int d\xi^-_N \hat q_F(\xi_N).
\label{eq:Delta2F}
\end{equation}
is given by the quark transport parameter,
\begin{equation}
\hat q_F(\xi_N)
=\frac{2\pi^2\alpha_s}{N_c}\rho_N^A(\xi_N)[xf^N_g(x)]_{x=0},
\label{qhat1}
\end{equation}
where $\rho_N^A(\xi_N)$ is the spatial nucleon number density inside the nucleus
and $f^N_g(x)$ is the gluon distribution function in a nucleon,
\begin{equation}
xf^N_g(x)=-\int\frac{d\xi^-}{2\pi p^+} e^{ixp^+\xi^-}
\langle N\mid F_{+\sigma}(0)F_+^{\;\;\sigma}(\xi^-)\mid N\rangle.
\end{equation}
A summation over the gluon's color index is implied in the above definition of the gluon distribution function.

One can generalize the above approach to nuclear modification of higher-twist TMD parton distributions.
In this paper, we will study the case of twist-3 TMD quark distributions which determines the azimuthal
asymmetry $\langle\cos\phi\rangle$ of the unpolarized semi-inclusive deep-inelastic scattering (SIDIS) cross
section defined with respect to the leptonic plane. Such asymmetry at large transverse momentum
arises predominately from hard gluon bremsstrahlung in perturbative QCD \cite{Georgi:1977tv,Bacchetta:2008xw}. Similar
processes are also responsible for the azimuthal angle dependence of the Drell-Yan dilepton production
cross sections \cite{Lam:1978pu} and multiple parton scattering in $p+A$ collisions
can lead to a twist-4 nuclear dependence \cite{Fries:1999jj,Fries:2000da}.
In the small transverse momentum $k_\perp \le 1$ GeV/c region, however, the asymmetry is shown to arise mainly
from the intrinsic transverse momentum of the quarks in the target
nucleon or nucleus \cite{Cahn:1978se,Berger:1979kz} and is related to the higher-twist TMD parton
distribution functions \cite{Mulders:1995dh,Oganesian:1997jq,Chay:1997qy,Liang:2006wp}. We will focus
on the twist-3 contribution to the azimuthal dependence,  proportional to $(k_\perp/Q)\cos\phi$,  in this
paper. Here $Q^2=-q^2$ and $q$ is the four momentum transfer from lepton to target.
Since the coefficient of the azimuthal asymmetry is related to twist-3 TMD parton distributions, one can take the same
approach as  in the case of the twist-2 TMD parton distribution to study the nuclear dependence of the
twist-3 contribution to the azimuthal asymmetry of SIDIS cross sections.



The remainder of this paper is organized as follows. In Sec. II, we will review the calculation of the
azimuthal dependence of SIDIS up to twist-3 contributions in terms of the TMD parton distributions.
In Sec. III, we extend the study of nuclear dependence of TMD parton distributions to the case of other
TMD parton correlation functions, including twist-3. Using these results,  we calculate the general nuclear
dependence of  $\langle\cos\phi\rangle$. We will then illustrate the nuclear dependence with
an ansatz of the TMD parton  distributions in a Gaussian form and discuss the effect
of transverse momentum  smearing in the fragmentation.  A summary will be given in Sec. IV.

\section{Azimuthal asymmetry in SIDIS}

We consider semi-inclusive deep inelastic scattering $e^-+A\to e^-+q+X$
with unpolarized beam and nucleus or nucleon target. The differential cross section,

\begin{equation}
d\sigma=\frac{\alpha_{em}^2e_q^2}{sQ^4}L^{\mu\nu}(l,l')\frac{d^2W_{\mu\nu}}{d^2k'_\perp}
\frac{d^3l'd^2k'_\perp}{(2E_{l'})},
\end{equation}
can be expressed as a product of the leptonic tensor,
\begin{equation}
L^{\mu\nu}(l,l')
=4[l^\mu{l'}^\nu+l^\nu{l'}^\mu-(l\cdot l')g^{\mu\nu}],
\end{equation}
and the hadronic tensor,
\begin{eqnarray}
\frac{d^2W_{\mu\nu}}{d^2k'_\perp}&=&\int \frac{dk'_z}{(2\pi)^3 2E_{k'}}W_{\mu\nu}^{(si)}(q,p,k'); \\
W_{\mu\nu}^{(si)}(q,p,k')&=&
\frac{1}{2\pi}\sum_X \langle A| J_\mu(0)|k',X\rangle \langle k',X| J_\nu(0)|A\rangle \nonumber\\
&&\times (2\pi)^4\delta^4(p+q-k'-p_X),
\end{eqnarray}
where the superscript $(si)$ denotes that it is for SIDIS. Here $l$ and $l'$
are the four momenta of the incoming and outgoing leptons, respectively,
$p$ is the four momentum per nucleon of the incoming target $A$,
$q$ is the four momentum transfer,  and $k'$ is the four momentum of the outgoing quark.
We neglect the masses and use the light-cone coordinates.
The unit vectors are taken as,
$\bar n=(1,0,0,0)$, $n=(0,1,0,0)$, $n_{\perp 1}=(0,0,1,0)$,
$n_{\perp 2}=(0,0,0,1)$.
We choose the coordinate system such that,
$p=p^+\bar n$, $q=-x_Bp+nQ^2/(2x_Bp^+)$, and
$l_\perp=|\vec l_\perp|n_{\perp 1}$,
where $x_B=Q^2/2p\cdot q$ is the Bjorken variable and $y=p\cdot q/p\cdot l$.

Since the azimuthal asymmetry $\langle\cos \phi\rangle$ in the kinematic region $k_\perp\ll Q$ is a twist-3 effect,
we need  to calculate $d^2W_{\mu \nu}/d^2k_\perp$ up to the twist-3 level. The calculations
have been carried out in Ref.~\cite{Liang:2006wp} and we review it here for our later study of nuclear
dependence.

In the unpolarized SIDIS $e^-+A\to e^-+q+X$, the twist-2 contribution is independent of the direction
of $\vec k_\perp$ and is given by,
\begin{equation}
\left[\frac{d^2W_{\mu \nu}}{d^2k_\perp}\right]_{\rm Twist-2}=
H^{(0)}_{\mu\nu}(x_B)f_q^A(x_B,k_\perp),
\end{equation}
where the hard part is
\begin{equation}
H^{(0)}_{\mu\nu}(x_B)=
\frac{1}{4p\cdot q}\textrm{Tr}\left[p\hspace{-6pt}\slash\gamma_{\mu}
(x_B p\hspace{-6pt}\slash+q \hspace{-6pt}\slash)\gamma_{\nu}\right]=-d_{\mu\nu},
\end{equation}
with $d^{\mu\nu}=g^{\mu\nu}-\bar{n}^{\mu}n^{\nu}-\bar{n}^{\nu}n^{\mu}$  as a projection tensor,
and $f_q^A(x,k_\perp)$ is TMD quark distribution function in the nucleus or nucleon,
\begin{widetext}
\begin{equation}
f_q^A(x,k_\perp)=\int \frac{dy^- d^2\vec{y}_{\perp}}{(2\pi)^3}
e^{ix p^+ y^- -i\vec{k}_{\perp}\cdot \vec{y}_{\perp}}
\langle A|\bar{\psi}(0)\frac{\gamma^+}{2}{\cal{L}}(0,y)\psi(y)|A\rangle,
\label{eq:fqndef}
\end{equation}
For brevity, we now work in the covariant gauge and use $y$ to denote four vector of coordinates $(0,y^-,\vec{y}_{\perp})$.
The gauge link ${\cal{L}}(0,y)$ is given by,
\begin{eqnarray}
\label{CollinearGL}
{\cal{L}}(0,y)&=&\mathcal{L}^\dag(-\infty,\vec{0}_\perp; 0,\vec{0}_\perp)
\mathcal{L}(-\infty,\vec{y}_{\perp}; y^-,\vec{y}_{\perp});
\end{eqnarray}
\begin{equation}
\label{TMDGL}
\mathcal{L}( -\infty,\vec{y}_{\perp};y^-,\vec{y}_{\perp})=
P \exp \left( i g \int^{y^-}_{-\infty} d \xi^{-} A^+ ( \xi^-, \vec{y}_{\perp})\right) .
\end{equation}

Twist-3 contributions to the semi-inclusive hadronic tensor include 3 terms which can result in  $\cos\phi$ azimuthal dependence,
\begin{equation}
\left[\frac{d^2W_{\mu \nu}}{d^2{k}_\perp}\right]_{\rm Twist-3}=
\left[\frac{d^2\tilde W^{(0)}_{\mu \nu}}{d^2{k}_\perp}\right]_{\rm Twist-3}+
\left[\frac{d^2\tilde W^{(1,L)}_{\mu \nu}}{d^2{k}_\perp}\right]_{\rm Twist-3}+
\left[\frac{d^2\tilde W^{(1,R)}_{\mu \nu}}{d^2{k}_\perp}\right]_{\rm Twist-3},
\end{equation}
where the number in the superscript of $\tilde W_{\mu\nu}$ denotes the number of gluon(s)
involved in the multiple gluon scattering with respect to which the collinear expansion is carried out;
the tilde above $W$ denotes the results after collinear expansion; and the superscript $L$ or $R$ denotes
the left or right cut respectively with respect to the initial gluon line.
The contribution $\tilde W^{(0)}$ is given by,
\begin{equation}
\left[\frac{d^2\tilde W^{(0)}_{\mu \nu}}{d^2k_\perp}\right]_{\rm Twist-3}=
H^{(0)}_{\mu\nu,\rho}(x_B) \Phi^{(0)A}_\sigma(x_B,\vec k_\perp)d^{\rho\sigma},
\end{equation}
where the hard part is given by,
\begin{equation}
H_{\mu\nu,\rho}^{(0)}(x)=\frac{1}{4p\cdot q} \textrm{Tr} \left[\gamma_\rho\gamma_\mu
(xp\hspace{-6pt}\slash+q\hspace{-6pt}\slash)\gamma_\nu \right],
\end{equation}
and the TMD parton correlation is defined as,
\begin{equation}
{\Phi}^{(0)A}_{\sigma}(x,\vec{k}_{\perp})
=\int \frac{p^+ dy^- d^{2}y_{\perp}}{(2\pi)^3}
e^{ix p^+ y^- -i\vec{k}_{\perp}\cdot \vec{y}_{\perp}}
\langle A|\bar{\psi}(0)\frac{\gamma_\sigma}{2}{\cal{L}}(0,y)\psi(y)|A\rangle.
\label{eq:Phi0}
\end{equation}
After carrying out the trace, we obtain,
\begin{equation}
\left[\frac{d^2\tilde W^{(0)}_{\mu \nu}}{d^2k_\perp}\right]_{\rm Twist-3}
=(k_{\perp\mu} n_\nu+k_{\perp\nu} n_\mu)f_{q\perp}^A(x_B,k_\perp),
\end{equation}
where the twist-3 parton distribution $f_{q\perp}^A(x_B,k_\perp)$ is defined as
\begin{equation}
k_\perp^\alpha f_{q\perp}^A(x_B,k_\perp)=d^{\alpha\beta}\Phi^{(0)A}_\beta(x,\vec k_\perp).
\label{eq:fqperp}
\end{equation}


The contribution $\tilde W^{(1,R)}$ and $\tilde W^{(1,L)}$ are the same in 
unpolarized SIDIS and are given by,
\begin{equation}
\frac{d^2\tilde W_{\mu\nu}^{(1,R)}}{d^2k_\perp}=
\frac{d^2\tilde W_{\mu\nu}^{(1,L)}}{d^2k_\perp}= \frac{1}{4p\cdot q}
{\rm Tr}[\hat{h}^{(1)\rho}_{\mu\nu}\omega_\rho^{\ \rho'}\hat\varphi^{(1)A}_{\rho'}(x_B,\vec k_\perp)],
\end{equation}
where $\omega_\rho^{\ \rho'}\equiv g_\rho^{\ \rho'}-\bar n_\rho n^{\rho'}$,
$\hat{h}^{(1)\rho}_{\mu\nu}=\gamma_\mu n\hspace{-6pt}\slash\gamma^\rho\bar{n}\hspace{-6pt}\slash\gamma_\nu$ 
and the matrix element
\begin{equation}
\hat\varphi^{(1)A}_\rho(x,\vec k_\perp)
=\int \frac{p^+ dy^-d^2y_\perp}{(2\pi)^3} e^{ixp^+y^--i\vec y_\perp\cdot\vec k_\perp}
\langle A|\bar\psi(0) {\cal L}(0,y)D_\rho(y)\psi(y)|A\rangle,
\label{eq:corr2}
\end{equation}
has two independent terms contributing to the hadronic tensor
\begin{equation}
\left[\frac{d^2\tilde W_{\mu\nu}^{(1)}}{d^2k_\perp}\right]_{\rm Twist-3}
=\frac{1}{4p\cdot q} \Bigl\{
{\rm Tr}[p\hspace{-5pt}\slash\hat{h}^{(1)\rho}_{\mu\nu}]k_{\perp\rho}\varphi^{(1)A}_{\perp}(x_B,k_\perp)
+i{\rm Tr}[\gamma_5p\hspace{-5pt}\slash\hat h^{(1)\rho}_{\mu\nu}]
\epsilon_{\perp\rho\gamma}k_{\perp}^\gamma\tilde\varphi^{(1)A}_{\perp}(x_B,k_\perp)\Bigr\},
\end{equation}
where $\epsilon_{\perp\rho\gamma}\equiv\epsilon_{\alpha\beta\rho\gamma}\bar n^\alpha n^\beta$.
After carrying out the traces and make the Lorentz contraction with $k_{\perp\rho}$
or $\epsilon_{\perp\rho\gamma}k_{\perp}^\gamma$ respectively, we obtain,
\begin{equation}
\left[\frac{d^2\tilde W_{\mu\nu}^{(1)}}{d^2k_\perp}\right]_{\rm Twist-3}
=-\frac{1}{p\cdot q}(p_\mu k_{\perp\nu}+p_\nu k_{\perp\mu})
[\varphi^{(1)A}_{\perp}(x_B,k_\perp)-\tilde\varphi^{(1)A}_{\perp}(x_B,k_\perp)],
\label{eq:W1Lres}
\end{equation}
where the parton correlation functions are given by,
\begin{equation}
k_\perp^2\varphi^{(1)A}_{\perp}(x,k_\perp)=n^\alpha k_\perp^\rho\varphi^{(1)A}_{\rho\alpha}(x,\vec{k}_\perp),
\label{eq:varphi1}
\end{equation}
\begin{equation}
ik_\perp^2\tilde\varphi^{(1)A}_{\perp}(x,k_\perp)=
n^\alpha \epsilon_{\perp}^{\rho\beta}k_{\perp\beta}\tilde\varphi^{(1)A}_{\rho\alpha}(x,\vec{k}_\perp),
\end{equation}

\begin{equation}
\varphi^{(1)A}_{\rho\alpha}(x,\vec k_\perp)=\int \frac{dy^-d^2y_\perp}{(2\pi)^3}e^{ixp^+y^--i\vec y_\perp\cdot\vec k_\perp}
\langle A|\bar\psi(0)\frac{\gamma_\alpha}{2}{\cal L}(0,y)D_\rho(y)\psi(y)|A\rangle
\label{eq:varphi2}
\end{equation}
\begin{equation}
\tilde\varphi^{(1)A}_{\rho\alpha}(x,\vec k_\perp)=\int \frac{dy^-d^2y_\perp}{(2\pi)^3}e^{ixp^+y^--i\vec y_\perp\cdot\vec k_\perp}
\langle A|\bar\psi(0)\frac{\gamma_5\gamma_\alpha}{2}{\cal L}(0,y)D_\rho(y)\psi(y)|A\rangle
\end{equation}

Equation of motion relates
\begin{equation}
xf_{q\perp}^A(x,k_\perp)=-\varphi_\perp^{(1)A}(x,k_\perp)+\tilde\varphi_\perp^{(1)A}(x,k_\perp),
\end{equation}
so we have,
\begin{equation}
\left[\frac{d^2\tilde W_{\mu\nu}}{d^2k_\perp}\right]_{\rm Twist-3}=
\left[\frac{d^2\tilde W_{\mu\nu}^{(0)}}{d^2k_\perp}+
\frac{d^2\tilde W_{\mu\nu}^{(1)}}{d^2k_\perp}\right]_{\rm Twist-3}
=\frac{1}{p\cdot q}[(q_\mu+2x_Bp_\mu) k_{\perp\nu}+(q_\nu+2x_Bp_\nu) k_{\perp\mu}]f^A_{q\perp}(x_B,k_\perp),
\label{eq:Wt3final}
\end{equation}

Summing both twist-2 and twist-3 contributions and contracting with lepton tensor $L_{\mu\nu}$, we obtain
the differential cross section as,
\begin{equation}
\frac{d\sigma}{dx_Bdyd^2k_\perp}=\frac{2\pi \alpha_{em}^2e_q^2}{Q^2y}
\Bigl\{ [1+(1-y)^2] f_q^A(x_B,\vec k_\perp)
- 4(2-y)\sqrt{1-y} \frac{|\vec k_\perp|}{Q} x_Bf_{q\perp}^{(1)A}(x_B,\vec k_\perp)\cos\phi \Bigr\},
\end{equation}
where the azimuthal angle $\phi$ is defined by $\cos\phi=\hat{\vec{k}}_\perp\cdot \hat{\vec{l}}_\perp$.
The azimuthal asymmetry at fixed $k_\perp$ is given by,
\begin{equation}
\langle \cos\phi\rangle_{eA}=- \frac{2(2-y)\sqrt{1-y}}{1+(1-y)^2}\frac{|\vec k_\perp|}{Q}
\frac{x_Bf_{q\perp}^A(x_B,k_\perp)}{f_q^A(x_B,k_\perp)}.
\end{equation}
We can also calculate the transverse-momentum integrated asymmetry and obtain,
\begin{equation}
\langle\langle \cos\phi\rangle\rangle_{eA}=- \frac{2(2-y)\sqrt{1-y}}{1+(1-y)^2}
\frac{\int |\vec k_\perp|d^2k_\perp x_Bf_{q\perp}^A(x_B,k_\perp)}{f_q^A(x_B)},
\end{equation}
where $f_q^A(x)=\int d^2k_\perp f_q^A(x,k_\perp)$ is the usual quark distribution in a nucleon or nucleus.

If we consider only ``free partons '' with intrinsic transverse momentum, {\em i.e.}, setting $g=0$
then, ${\cal L}=1$ , $D_{\sigma}=\partial_{\sigma}$, and $xf_{q\perp}^A(x,k_\perp)=f_q^A(x,k_\perp)$. In this case,
$\langle \cos\phi\rangle_{eA}=-2(2-y)\sqrt{1-y}/[1+(1-y)^2]\cdot |\vec k_\perp|/Q$, which is just the
result obtained in Ref.~\cite{Cahn:1978se}.

We note that the above calculations apply to SIDIS of both nuclear and nucleon targets.
All results are in the same form in terms of the TMD parton distributions inside a nucleus or
a nucleon. The nuclear dependence of the azimuthal angle asymmetry will come from the
nuclear dependence of the TMD parton distributions. In the SIDIS off a nuclear target,
multiple gluon scattering between the struck quark and the target as contained in the gauge
links can happen to different nucleons inside the nucleus. This will give rise to the nuclear dependence
which we will discuss in the remainder of this paper.


\section{$A$-dependence of the azimuthal asymmetry}

Following the same approach in the discussion of the  nuclear dependence of the twist-2 TMD
quark distribution $f_q^A(x,k_\perp)$ in Ref.~\cite{Liang:2008vz}, we can also express
the general parton distribution,
\begin{eqnarray}
\Phi^{A}_\alpha(x,k_\perp) & \equiv& \int \frac{p^+ dy^-}{2\pi} \frac{d^2y_\perp}{(2\pi)^2}
e^{ixp^+y^- -i\vec  k_\perp\cdot \vec y_\perp}
\langle A \mid \bar\psi(0)\frac{\Gamma_\alpha }{2}{\cal L}(0,y)\psi(y)\mid A \rangle,
\label{form} \\
&=&\int \frac{p^{+}dy^-}{2\pi}e^{ixp^+y^-}
\langle A \mid \bar\psi(0)\frac{\Gamma_\alpha}{2}
e^{\vec W_\perp(y^-) \cdot\vec\nabla_{ k_\perp}}\psi(y^-)\mid A \rangle
\delta^{(2)}(\vec k_\perp),
\label{tmd2}
\end{eqnarray}
in terms of the collinear parton matrix elements involving the transport operator
$\vec W_\perp(y^-)$ [Eq.~(\ref{eq:transop})],
where $\Gamma_\alpha$ is any gamma matrix. Expanding the exponential term of the above
matrix element in powers of the transport operator $\vec W_{\perp}(y^{-})$ and assuming
``maximum two-gluon correlation approximation'' as in Ref.~\cite{Liang:2008vz}, one can
express the nuclear TMD parton distributions in terms of a Gaussian convolution of the same TMD distributions in a nucleon,
\begin{eqnarray}
\Phi^{A}_\alpha(x,k_\perp)
& \approx & A \exp\left[\frac{\Delta_{2F}}{4}\nabla_{ k_\perp}^2\right] \Phi^{N}_\alpha(x,k_\perp),
\label{tmd4} \\
&=&\frac{A}{\pi \Delta_{2F}}
\int d^2\ell_\perp e^{-(\vec k_\perp -\vec\ell_\perp)^2/\Delta_{2F}}\Phi^{N}_\alpha(x,\ell_\perp).
\label{tmdgeneral}
\end{eqnarray}
Note that in the derivation of the above result, one has considered the fact that matrix elements with
odd powers of $\vec W_{\perp}(y^{-})$ vanish for $\Gamma_{\alpha}=\gamma_{+}$ while the matrix
elements with even powers of $\vec W_{\perp}(y^{-})$ vanish for $\Gamma_{\alpha} =\vec\gamma_{\perp}$.

The nuclear broadening of the twist-2 TMD quark distribution is a special case of the above general result
with $\Gamma^{\alpha}=\gamma^{+}$. With the definition of the twist-3 parton distribution
function $f^A_{q\perp}(x,k_\perp)$ in Eqs.~(\ref{eq:Phi0}) and (\ref{eq:fqperp}), one can multiply both sides
of Eq.~(\ref{tmdgeneral}) by $d^{\alpha\beta}$ and obtain,
\begin{eqnarray}
f_{q\perp}^A(x,k_\perp)& \approx &
\frac{A}{\pi \Delta_{2F}}
\int d^2\ell_\perp \frac{(\vec k_\perp\cdot\vec\ell_\perp)}{\vec k_\perp^{2}}
e^{-(\vec k_\perp -\vec\ell_\perp)^2/\Delta_{2F}}f_{q\perp}^N(x,\ell_\perp) \\
&=&\frac{A}{\pi \Delta_{2F}}
\left(1+ \frac{\Delta_{2F}}{2\vec k_{\perp}^2}
\vec k_{\perp}\cdot\vec\partial _{k_{\perp}}\right)
\int d^2\ell_\perp e^{-(\vec k_\perp -\vec\ell_\perp)^2/\Delta_{2F}}f_{q\perp}^N(x,\ell_\perp).
\label{eq:fqperp2}
\end{eqnarray}

Given the twist-2 TMD quark distribution function $f_{q}^N(x, k_\perp)$ and the twist-3  quark distribution
$f_{q\perp}^N(x,\ell_\perp)$ in a nucleon and using the above convolution, Eqs.(\ref{eq:tmdNA}) and 
(\ref{eq:fqperp2}), one can then calculate the nuclear dependence of the azimuthal asymmetry
$\langle \cos\phi\rangle$ and   $\langle\langle \cos\phi\rangle\rangle$.
In general, if the convoluted  twist-3 TMD quark
distribution is a decreasing function of the transverse momentum in the region of interest, the second term involving a
derivative in the above equation will be negative. Therefore, one would expect the azimuthal asymmetry to decrease
because of the multiple scattering in nuclei.

To illustrate the nuclear dependence of the azimuthal asymmetry qualitatively, we consider an ansatz of
the Gaussian distributions in $k_\perp$ for both the twist-2 and 3 TMD quark distributions,
\begin{equation}
f_q^N(x,k_\perp)=\frac{1}{\pi\alpha}f_q^N(x)e^{- k_\perp^2/\alpha},
\end{equation}
\begin{equation}
f_{q\perp}^N(x,k_\perp)=\frac{1}{\pi\beta}f_{q\perp}^N(x)e^{- k_\perp^2/\beta}.
\end{equation}
The corresponding TMD distributions in nuclei are,
\begin{equation}
f_q^A(x,k_\perp)\approx\frac{A}{\pi(\alpha+\Delta_{2F})}f_q^N(x)e^{- k_\perp^2/(\alpha+\Delta_{2F})},
\end{equation}
\begin{equation}
f_{q\perp}^A(x,k_\perp)\approx\frac{A\beta}{\pi(\beta+\Delta_{2F})^2}
f_{q\perp}^N(x)e^{- k_\perp^2/(\beta+\Delta_{2F})}.
\end{equation}

One can then calculate the the azimuthal asymmetry for SIDIS off both nucleon
\begin{equation}
\langle \cos\phi\rangle_{eN}=- \frac{2(2-y)\sqrt{1-y}}{1+(1-y)^2}\frac{\alpha}{\beta}\frac{|\vec k_\perp|}{Q}
\frac{x_Bf_{q\perp}^N(x_B)}{f_q^N(x_B)}\exp\Bigl\{-\frac{\alpha-\beta}{\alpha\beta} k_\perp^2\Bigr\},
\label{eq:cosphieN}
\end{equation}
and nuclear targets
\begin{equation}
\langle \cos\phi\rangle_{eA}=-\frac{2(2-y)\sqrt{1-y}}{1+(1-y)^2}\frac{\beta(\alpha+\Delta_{2F})}{(\beta+\Delta_{2F})^2}
\frac{|\vec k_\perp|}{Q}
\frac{x_Bf_{q\perp}^N(x)}{f_q^N(x)}
\exp\Bigl\{-\frac{\alpha-\beta}{(\alpha+\Delta_{2F})(\beta+\Delta_{2F})} k_\perp^2\Bigr\}.
\label{eq:cosphieA}
\end{equation}
The nuclear modification factor for the azimuthal asymmetry is then,
\begin{equation}
\frac{\langle \cos\phi\rangle_{eA}}{\langle \cos\phi\rangle_{eN}}
=\frac{\beta^2(\alpha+\Delta_{2F})}{\alpha(\beta+\Delta_{2F})^2}
\exp\Bigl\{\frac{(\alpha-\beta)\Delta_{2F}(\alpha+\beta+\Delta_{2F})}
{\alpha\beta(\alpha+\Delta_{2F})(\beta+\Delta_{2F})}\vec k_\perp^2\Bigr\}.
\end{equation}

In the special case when $\alpha=\beta$, we have
\begin{equation}
\langle \cos\phi\rangle_{eN}=- \frac{2(2-y)\sqrt{1-y}}{1+(1-y)^2}\frac{|\vec k_\perp|}{Q}
\frac{x_Bf_{q\perp}^N(x_B)}{f_q^N(x_B)},
\label{eq:cosphieN2}
\end{equation}
\begin{equation}
\langle \cos\phi\rangle_{eA}=- \frac{2(2-y)\sqrt{1-y}}{1+(1-y)^2}\frac{\beta}{\beta+\Delta_{2F}}
\frac{|\vec k_\perp|}{Q}\frac{x_Bf_{q\perp}^N(x)}{f_q^N(x)},
\label{eq:cosphieA2}
\end{equation}
\begin{equation}
\frac{\langle \cos\varphi\rangle_{eA}}{\langle \cos\varphi\rangle_{eN}}
=\frac{\beta}{\beta+\Delta_{2F}}.
\end{equation}
Therefore, the azimuthal asymmetry $\langle\cos\phi\rangle_{eA}$ in deep inelastic $eA$ scattering
is suppressed compared to that in $eN$ scattering and the suppression is inversely proportional to
the total transverse momentum broadening $\Delta_{2F}$. The suppression is independent
of the transverse momentum $k_{\perp}$. However, in general, the twist-2 and twist-3 TMD quark distributions are not necessarily
the same and their Gaussian ansatz might have different widths $\beta \neq \alpha$. The nuclear modification factor for the
azimuthal asymmetry will then have non-trivial $k_{\perp}$ dependence. Shown in Fig. 1 are the nuclear modification
factors for the azimuthal asymmetry when $\beta/\alpha=2$ and 0.5, respectively, as functions of $\Delta_{2F}/\alpha$,
at different transverse momentum $k_{\perp}$. In the case $\beta>\alpha$, we see that the azimuthal asymmetry is suppressed
and the suppression increases with the transverse momentum $k_{\perp}$. However, when $\beta<\alpha$, the suppression
actually decreases with increasing $k_{\perp}$ and the azimuthal asymmetry could be enhanced for large enough
transverse momentum $k_{\perp}$. Therefore, the nuclear modification of the azimuthal asymmetry and its transverse
momentum dependence is a very sensitive probe of the twist-2 and twist-3 TMD quark distribution functions.

   \begin{figure} [htb]
    \resizebox{0.6\textwidth}{!}{\includegraphics{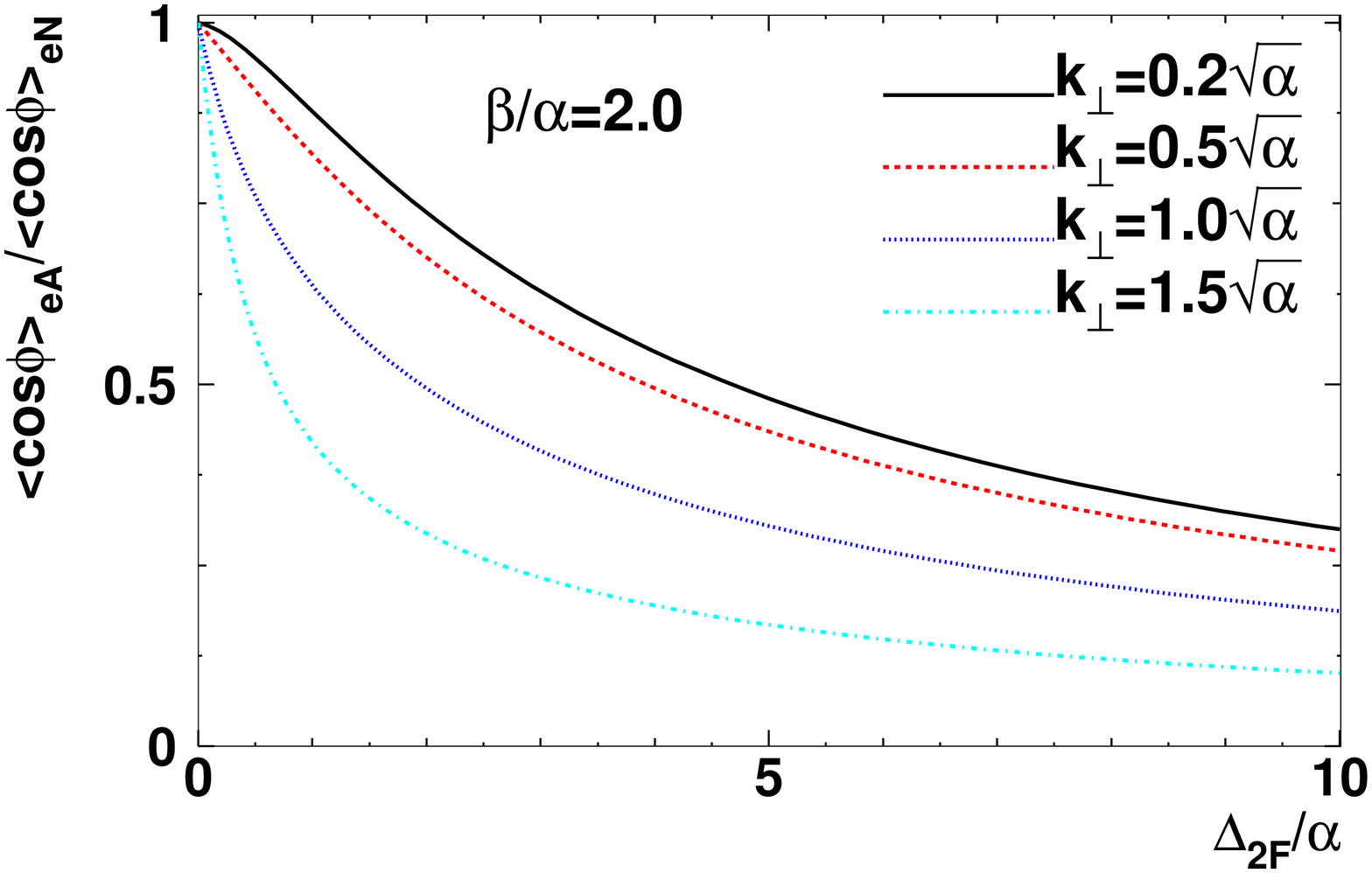}}
    \resizebox{0.6\textwidth}{!}{\includegraphics{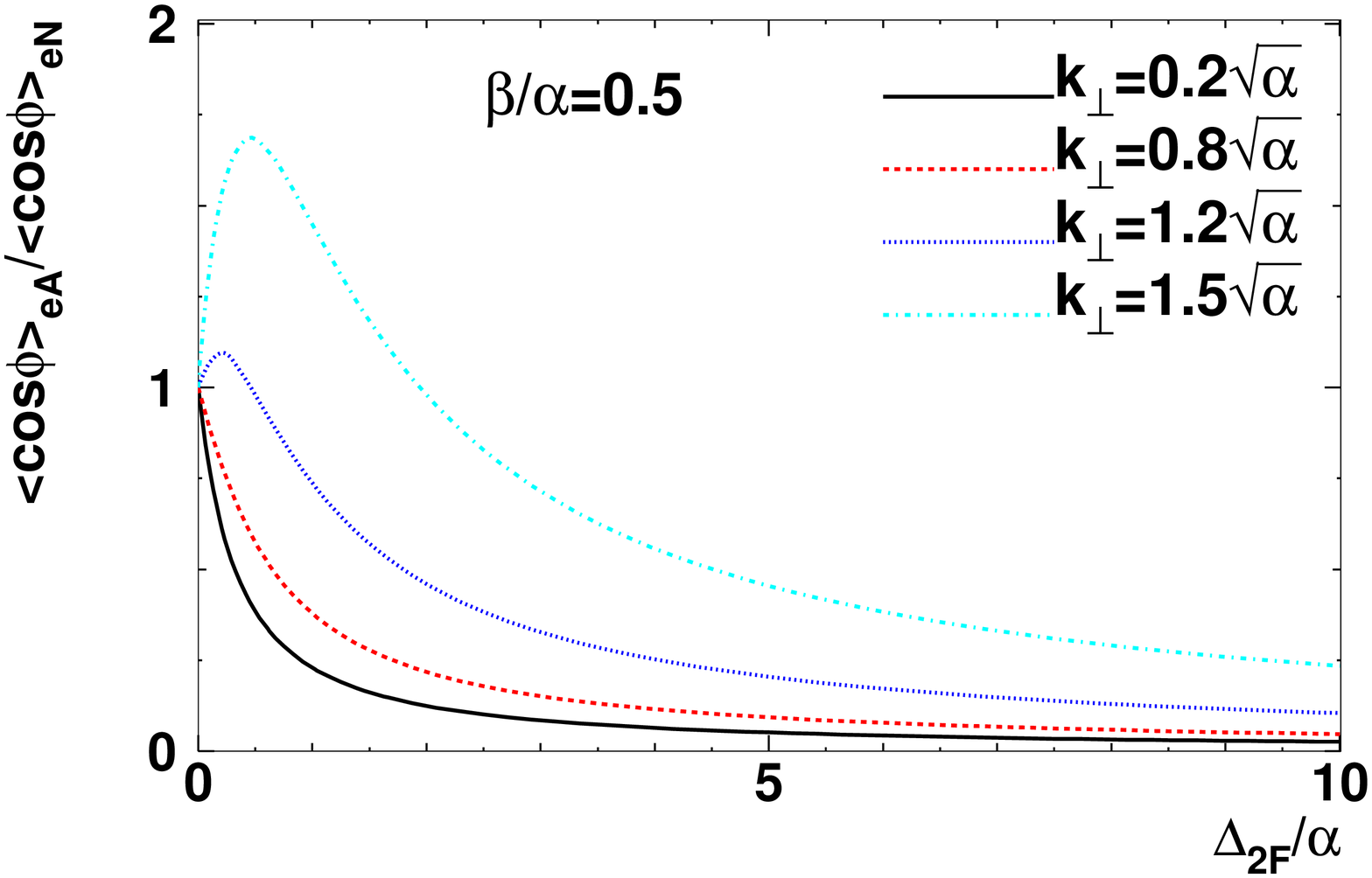}}
    \caption{(color online) Ratio $\langle \cos\phi\rangle_{eA}/\langle \cos\phi\rangle_{eN}$ as a
    function of $\Delta_{2F}$ for different $k_\perp$ and the relative width $\beta/\alpha$ of twist-3 and 2
    TMD quark distributions.  }
    \label{fig:AfH}       
    \end{figure}

If we integrate over the magnitude of the transverse momentum $k_\perp$, the averaged azimuthal asymmetry will only depend on the
shape of the twist-3 TMD quark distributions,
\begin{equation}
\langle\langle \cos\phi\rangle\rangle_{eN}=-\frac{2(2-y)\sqrt{1-y}}{1+(1-y)^2}
\frac{\sqrt{\pi\beta}}{2Q}
\frac{x_Bf_{q\perp}^N(x)}{f_q^N(x)},
\end{equation}
\begin{equation}
\langle\langle \cos\phi\rangle\rangle_{eA}=- \frac{2(2-y)\sqrt{1-y}}{1+(1-y)^2}
\frac{\sqrt{\pi}\beta}{2Q\sqrt{\beta+\Delta_{2F}}}
\frac{x_Bf_{q\perp}^N(x)}{f_q^N(x)},
\end{equation}
\begin{equation}
\frac{\langle\langle \cos\phi\rangle\rangle_{eA}}{\langle\langle \cos\phi\rangle\rangle_{eN}}
=\sqrt{\frac{\beta}{\beta+\Delta_{2F}}}.
\end{equation}
We see again that $\langle\langle\cos\phi\rangle\rangle_{eA}$
is suppressed compared to$\langle\langle\cos\phi\rangle\rangle_{eN}$,
and the suppression factor is inversely proportional to the square-root of the total transverse
momentum broadening $\Delta_{2F}$.

To take into account of the transverse momentum during the quark  fragmentation process and its
effect on the azimuthal asymmetry, we take another Gaussian smearing
\begin{equation}
d\sigma^{eN\to ehX}=\int d\sigma^{eN\to eqX}D_F^{q\to h}(z,\vec k_{F\perp})dzd^2k_{F\perp}
\delta^{(3)}(\vec p_h-z\vec k'-k_{F\perp})d^3p_h,
\end{equation}
for the $k_{F\perp}$-dependence in the fragmentation function $D_F^{q\to h}(z,\vec k_{F\perp})$, {\it i.e.},
\begin{equation}
D_F^{q\to h}(z,\vec k_{F\perp})=D_F^{q\to h}(z)\frac{1}{\pi\alpha_F}e^{-\vec k_{F\perp}^2/\alpha_F}.
\end{equation}
Consider the case that $\alpha=\beta$ and one has the azimuthal asymmetry for the hadron production cross section
\begin{equation}
\langle \cos\phi_h\rangle_{eN}=-\frac{2(2-y)\sqrt{1-y}}{1+(1-y)^2}\frac{\beta z}{\beta z^2+\alpha_F}
\frac{|\vec p_{h\perp}|}{Q}
\frac{x_Bf_{q\perp}^N(x)}{f_q^N(x)},
\end{equation}
\begin{equation}
\langle \cos\phi_h\rangle_{eA}=- \frac{2(2-y)\sqrt{1-y}}{1+(1-y)^2}
\frac{\beta z}{(\beta+\Delta_{2F})z^2+\alpha_F}
\frac{|\vec p_{h\perp}|}{Q}
\frac{x_Bf_{q\perp}^N(x)}{f_q^N(x)}.
\end{equation}
We compare the above results with those obtained without fragmentation, Eqs. (\ref{eq:cosphieN2}) and (\ref{eq:cosphieA2}),
and see that we have a clear smearing effect on the azimuthal asymmetry in both $e^-N$ and $e^-A$-scatterings.
The smearing factors are given by,
\begin{equation}
\frac{\langle\cos\phi_h\rangle_{eN}}{\langle \cos\phi\rangle_{eN}}\Bigr|_{|\vec p_{h\perp}|=z|\vec k_\perp|}
=\frac{\beta z^2}{\beta z^2+\alpha_F},
\end{equation}
\begin{equation}
\frac{\langle \cos\phi_h\rangle_{eA}}{\langle \cos\phi\rangle_{eA}}\Bigr|_{|\vec p_{h\perp}|=z|\vec k_\perp|}
=\frac{(\beta+\Delta_{2F}) z^2}{(\beta+\Delta_{2F}) z^2+\alpha_F}.
\end{equation}

The suppression factor in $eA$ compared to $eN$ is given by,
\begin{equation}
\frac{\langle \cos\phi_h\rangle_{eA}}{\langle \cos\phi_h\rangle_{eN}}
=\frac{\beta z^2+\alpha_F}{(\beta+\Delta_{2F})z^2+\alpha_F}.
\end{equation}
After integrated over the magnitude of the transverse momentum, we have,
\begin{equation}
\frac{\langle\langle \cos\phi_h\rangle\rangle_{eA}}{\langle\langle \cos\phi_h\rangle\rangle_{eN}}
=\sqrt{\frac{\beta z^2+\alpha_F}{(\beta+\Delta_{2F})z^2+\alpha_F}}.
\end{equation}

\section{Summary and discussions}
Within the generalized factorization, we have calculated the SIDIS cross sections in
terms of the TMD quark distributions in a nucleon or nucleus up to twist-3. The azimuthal
asymmetry $\langle \cos\phi\rangle$ in the small transverse momentum region depends on both
twist-2 and 3 TMD quark distributions. By considering nuclear broadening of both twist-2 and 3 TMD
quark distributions due to multiple scattering between the struck quark and nucleons inside the nucleus,
we investigated the nuclear dependence of the azimuthal asymmetry. We found that the azimuthal
asymmetry is suppressed by multiple parton scattering for most cases of the TMD quark distributions.
The suppression is inversely proportional to the average squared transverse momentum broadening.
The transverse momentum dependence of the suppression depends on the relative shape of the twist-2
and 3 TDM quark distributions. Using a Gaussian ansatz, we found that the suppression factor decreases with the
transverse momentum if the width of the twist-2 TMD distribution is smaller than that of the twist-3 TMD
distribution, while the suppression factor increases with the transverse momentum if the width of the twist-3
TMD distribution is smaller than that of the twist-2 TMD distribution. The suppression is independent
of the transverse momentum if the twist-2 and 3 TMD distributions have the same width. Therefore,
study of the nuclear dependence of the azimuthal asymmetry can shed light on the relative shape of
the twist-2 and 3 TMD quark distributions.

XNW would like to thank F. Yuan for help discussions. This work was supported in part by the National Natural 
Science Foundation of China under the project  Nos. 10525523 and 10975092,
the Department of Science and Technology of
Shandong Province and  the Director, Office of Energy
Research, Office of High Energy and Nuclear Physics, Division of
Nuclear Physics, of the U.S. Department of Energy under Contract No.
DE-AC02-05CH11231.

\end{widetext}


\begin{thebibliography}{}
\bibitem{VanHaarlem:2007kj}
  Y.~Van Haarlem, A.~Jgoun and P.~Di Nezza,
\bibitem{VanHaarlem:2009zz}
  Y.~Van Haarlem,
  Nucl.\ Phys.\ Proc.\ Suppl.\  {\bf 186}, 106 (2009).


\bibitem{Cronin:1974zm}
  J.~W.~Cronin, H.~J.~Frisch, M.~J.~Shochet, J.~P.~Boymond, R.~Mermod, P.~A.~Piroue and R.~L.~Sumner,
  Phys.\ Rev.\  D {\bf 11}, 3105 (1975).
\bibitem{Antreasyan:1978cw}
  D.~Antreasyan, J.~W.~Cronin, H.~J.~Frisch, M.~J.~Shochet, L.~Kluberg, P.~A.~Piroue and R.~L.~Sumner,
  Phys.\ Rev.\  D {\bf 19}, 764 (1979).



\bibitem{Gyulassy:1993hr}
  M.~Gyulassy and X.~N.~Wang,
  Nucl.\ Phys.\  B {\bf 420}, 583 (1994)

\bibitem{Baier:1996sk}
  R.~Baier, Y.~L.~Dokshitzer, A.~H.~Mueller, S.~Peigne and D.~Schiff,
  Nucl.\ Phys.\  B {\bf 484}, 265 (1997)

\bibitem{Wiedemann:2000za}
  U.~A.~Wiedemann,
  Nucl.\ Phys.\  B {\bf 588}, 303 (2000)

\bibitem{Gyulassy:2000er}
  M.~Gyulassy, P.~Levai and I.~Vitev,
  Nucl.\ Phys.\  B {\bf 594}, 371 (2001)

\bibitem{Guo:2000nz}
  X.~F.~Guo and X.~N.~Wang,
  Phys.\ Rev.\ Lett.\  {\bf 85}, 3591 (2000)

\bibitem{Wang:2001ifa}
  X.~N.~Wang and X.~F.~Guo,
  Nucl.\ Phys.\  A {\bf 696}, 788 (2001)


\bibitem{CasalderreySolana:2007sw}
  J.~Casalderrey-Solana and X.~N.~Wang,
  Phys.\ Rev.\  C {\bf 77}, 024902 (2008)



\bibitem{Adcox:2001jp}
  K.~Adcox {\it et al.}  [PHENIX Collaboration],
  Phys.\ Rev.\ Lett.\  {\bf 88}, 022301 (2002)

\bibitem{Adler:2002xw}
  C.~Adler {\it et al.}  [STAR Collaboration],
  Phys.\ Rev.\ Lett.\  {\bf 89}, 202301 (2002)

\bibitem{Adler:2002tq}
  C.~Adler {\it et al.}  [STAR Collaboration],
  Phys.\ Rev.\ Lett.\  {\bf 90}, 082302 (2003)


\bibitem{Wang:2002ri}
  E.~Wang and X.~N.~Wang,
  Phys.\ Rev.\ Lett.\  {\bf 89}, 162301 (2002)


\bibitem{Bass:2008rv}
  S.~A.~Bass, C.~Gale, A.~Majumder, C.~Nonaka, G.~Y.~Qin, T.~Renk and J.~Ruppert,
  Phys.\ Rev.\  C {\bf 79}, 024901 (2009)



\bibitem{Dolejsi:1993iw}
  J.~Dolejsi, J.~Hufner and B.~Z.~Kopeliovich,
  Phys.\ Lett.\  B {\bf 312}, 235 (1993)

\bibitem{Johnson:2000dm}
  M.~B.~Johnson, B.~Z.~Kopeliovich and A.~V.~Tarasov,
  Phys.\ Rev.\  C {\bf 63}, 035203 (2001)


\bibitem{Domdey:2008aq}
  S.~Domdey, D.~Grunewald, B.~Z.~Kopeliovich and H.~J.~Pirner,
  Nucl.\ Phys.\  A {\bf 825}, 200 (2009) 


 \bibitem{Luo:1992fz}
  M.~Luo, J.~W.~Qiu and G.~Sterman,
  Phys.\ Lett.\  B {\bf 279}, 377 (1992);
 %
  Phys.\ Rev.\  D {\bf 49}, 4493 (1994);
  %
  Phys.\ Rev.\  D {\bf 50}, 1951 (1994).


 \bibitem{Guo:1998rd}
  X.~F.~Guo,
  Phys.\ Rev.\  D {\bf 58}, 114033 (1998)


  \bibitem{Osborne:2002st}
  J.~Osborne and X.~N.~Wang,
  Nucl.\ Phys.\  A {\bf 710}, 281 (2002) 


\bibitem{Raufeisen:2003zk}
  J.~Raufeisen,
  Phys.\ Lett.\  B {\bf 557}, 184 (2003)

\bibitem{Liang:2008vz}
  Z.~T.~Liang, X.~N.~Wang and J.~Zhou,
  Phys.\ Rev.\  D {\bf 77}, 125010 (2008)


  \bibitem{Bodwin:1988fs}
  G.~T.~Bodwin, S.~J.~Brodsky and G.~P.~Lepage,
  Phys.\ Rev.\  D {\bf 39}, 3287 (1989).

  \bibitem{Michael:1981yy}
  C.~Michael and G.~Wilk,
  Z.\ Phys.\  C {\bf 10}, 169 (1981).


\bibitem{Guo:1999eh}
  X.~F.~Guo, J.~W.~Qiu and X.~F.~Zhang,
  Phys.\ Rev.\  D {\bf 62}, 054008 (2000)





\bibitem{Georgi:1977tv}
  H.~Georgi and H.~Politzer,
  Phys.\ Rev.\ Lett.\  {\bf 40}, 3 (1978).

\bibitem{Bacchetta:2008xw}
  A.~Bacchetta, D.~Boer, M.~Diehl and P.~J.~Mulders,
  JHEP {\bf 0808}, 023 (2008)
  [arXiv:0803.0227 [hep-ph]].



\bibitem{Lam:1978pu}
  C.~S.~Lam and W.~K.~Tung,
  Phys.\ Rev.\  D {\bf 18}, 2447 (1978).


\bibitem{Fries:1999jj}
  R.~J.~Fries, B.~Muller, A.~Schafer and E.~Stein,
  Phys.\ Rev.\ Lett.\  {\bf 83}, 4261 (1999)

\bibitem{Fries:2000da}
  R.~J.~Fries, A.~Schafer, E.~Stein and B.~Muller,
  Nucl.\ Phys.\  B {\bf 582}, 537 (2000)



\bibitem{Cahn:1978se}
  R.~N.~Cahn,
  Phys.\ Lett.\ B {\bf 78}, 269 (1978).

\bibitem{Berger:1979kz}
 E.~L.~Berger,
 Phys.\ Lett.\ B {\bf 89}, 241 (1980).


\bibitem{Mulders:1995dh}
  P.~J.~Mulders and R.~D.~Tangerman,
  Nucl.\ Phys.\ B {\bf 461}, 197 (1996)
  [Erratum {\bf 484}, 538 (1997)].

\bibitem{Oganesian:1997jq}
 K.~A.~Oganesian, H.~R.~Avakian, N.~Bianchi and P.~Di Nezza,
  Eur.\ Phys.\ J.\ C {\bf 5}, 681 (1998).

\bibitem{Chay:1997qy}
  J.~Chay and S.~M.~Kim,
  Phys.\ Rev.\ D {\bf 57}, 224 (1998)


\bibitem{Liang:2006wp}
  Z.~T.~Liang and X.~N.~Wang,
  Phys.\ Rev.\  D {\bf 75}, 094002 (2007)


\bibitem{Aubert:1983cz}
  J.~J.~Aubert {\it et al.}  [European Muon Collaboration],
  Phys.\ Lett.\ B {\bf 130}, 118 (1983);
%
\bibitem{Arneodo:1986cf}
  M.~Arneodo {\it et al.}  [European Muon Collaboration],
  Z.\ Phys.\ C {\bf 34}, 277 (1987).

\bibitem{Adams:1993hs}
  M.~R.~Adams {\it et al.}  [E665 Collaboration],
  Phys.\ Rev.\ D {\bf 48}, 5057 (1993).

\bibitem{Breitweg:2000qh}
  J.~Breitweg {\it et al.}  [ZEUS Collaboration],
  Phys.\ Lett.\ B {\bf 481}, 199 (2000);
%



\bibitem{Chekanov:2002sz}
  S.~Chekanov {\it et al.}  [ZEUS Collaboration],
  Phys.\ Lett.\ B {\bf 551}, 226 (2003).
%



\end{thebibliography}
\end{document}